\documentclass[preprint,aps,pre,floatfix]{revtex4-1}

\usepackage{graphicx}
\usepackage{epstopdf} 
\usepackage{dcolumn}
\usepackage{bm}
\usepackage{units}

\DeclareGraphicsExtensions{.pdf,.png,.jpg,.jpeg}

\newcommand{\bfE}{\mathbf{E}}
\newcommand{\bfB}{\mathbf{B}}

\newcommand{\bfxhat}{\mathbf{\hat{x}}}
\newcommand{\bfzhat}{\mathbf{\hat{z}}}
\newcommand{\bfJ}{\mathbf{J}}
\newcommand{\bfS}{\mathbf{S}}
\newcommand{\alf}{{Alfv\'en }}
\begin{document}

\title{Electromagnetic Energy Conversion  in Downstream Fronts from 3D Kinetic Reconnection}

\author{Giovanni Lapenta}
\affiliation{Departement Wiskunde, KU Leuven, Universiteit Leuven, Belgium}
\author{Martin Goldman, David Newman}
\affiliation{University of Colorado, USA}
\author{Stefano Markidis}
\affiliation{High Performance Computing and Visualization (HPCViz) Department, KTH Royal Institute of Technology,
Stockholm, Sweden}
\author{Andrey Divin}
\affiliation{Swedish Institute of Space Physics, Uppsala, Sweden}

\date{\today}

\begin{abstract}
The electromagnetic energy equation is analyzed term by term in a 3D simulation of kinetic reconnection previously reported by \citet{vapirev2013formation}. The evolution presents the usual 2D-like topological structures caused by an initial perturbation independent of the third dimension. However, downstream of the reconnection site, where the jetting plasma encounters the yet unperturbed pre-existing plasma, a downstream front (DF) is formed and made unstable by the strong density gradient and the unfavorable local acceleration field. The energy exchange between plasma and fields is most intense at the instability, reaching several $\unit{pW/m^3}$, alternating between load (energy going from fields to particles) and generator (energy going from particles to  fields) regions. Energy exchange is instead purely that of a load at the reconnection site itself in a region focused around the x-line and elongated along the separatrix surfaces. 
 Poynting fluxes are generated at all energy exchange regions and travel away from the reconnection site  transporting an energy signal of the order of about $\bfS \approx 10^{-3} \unit{W/m^2}$. 
\end{abstract}

\maketitle

\section{Introduction}

Reconnection is one of the most studied processes capable of releasing magnetic energy into kinetic energy, in the form of flows and particle heating. The typical conditions for reconnection have a sheared magnetic field where across an interface at least one component of the magnetic field reverses sign. In that situation the field lines of opposite polarity break and reconnect in a new configuration. In the process the magnetic energy content is decreased in favor of energization of the particles, in the form of ordered flows or random particle heating. 

The exact point where the magnetic field breaks needs not bear any direct link with the region where energy is in fact released. Traditionally, two competing scenarios have been guiding the discussion. In the Sweet-Parker mechanism \citep{sweet,parker} a diffusion region is present around the point of topological line breakage and the energy dissipation in the form of Ohmic heating and plasma acceleration is assumed to take place in the diffusion layer. In competition, the Petschek \citep{petschek} model sees standing slow shocks form an interface across which energy is released from the magnetic field to the particles. 

More recently, the kinetic description has identified that aspects of both models are present in the real plasma (see \citet{birn-priest} for a recent review). And real collisionless plasmas are kinetic of course. The electrons and ions become decoupled~\cite{sonnerup1979solar,terasawa1983hall} and two layers reminiscent of the Sweet-Parker layer are present \cite{laval1966instabilites,hesse1999diffusion,kuznetsova2000toward,gem1836}. An inner one has the electrons being accelerated and the outer one the ions. In the typical cartoon picture the two layers are nested boxes. Of course the cartoon is not real and the reality is more complex with the boxes extending along the separatrices, especially in presence of finite guide fields~\cite{kleva,biskamp,ricciguide}. 

But in kinetic reconnection, the energy release is not limited to the electron and ion diffusion regions proper. Other energy releases are possible in conduction with reconnection. A main process receiving substantial attention both from the theoretical simulation side and from direct in situ observation is that of the regions downstream of a reconnection site, where the plasma flow caused by reconnection slams into its surrounding plasma. In these regions the outflowing plasma and magnetic field act as a snowplow releasing its mass and energy against the surrounding pre-existing plasma and field. 

Such fronts emanating from a reconnection site have been studied in recent  kinetic simulations \citep{sitnov2009dipolarization} and confirmed by direct observational evidence has been obtained with data from the THEMIS mission~\citep{runov2009themis}. Many effects relative to these fronts can be studied in 2D, but one key process requires a full 3D study:  the presence of an instability that  perturbs the downstream front (DF)~\citep{nakamura2002interchange,guzdar2010simple}. Fluid studies have been used to capture the effect~\cite{nakamura2002interchange,guzdar2010simple}. In the fluid case, the instability squires the nature of an interchange mode similar to the Rayleigh-Taylor instability.  
 Across the downstream front the density increases substantially. The vertical magnetic field and the density pile up at the DF. The curvature of the field lines and the braking of the front by the momentum exchange with the yet unperturbed plasma leads to an effective acceleration pointing contrary to the DF speed. Such configuration is unstable to interchange modes: the higher density region is ahead of the front and the density gradient is therefore opposite to the direction of the acceleration~\citep{nakamura2002interchange}.

A well know limitation of MHD in this circumstance is the inability to predict the fastest growing mode: the interchange instability has the same growth speed at any wavenumber $k$. This is circumvented in MHD by seeding the instability ad hoc~\citep{guzdar2010simple} or self-contently as a consequence of other processes that produce the required seed~\citep{lapenta2011self}. 

Full kinetic studies are more suitable for modeling the process, but of course 3D kinetic simulations are much more computationally demanding, limiting the accessible domain size. In kinetic theory, extra physics is present to determine the scale of the process, predicting a fastest growing mode~\citep{pritchett2010kinetic}. In particular the presence of a density (and pressure) gradient at the DF induces also drift waves and instabilities that modify the nature of the DF instability~\citep{divin2013dipolarization}. Recently, \citet{vapirev2013formation} reported a fully kinetic simulation of the development of the secondary instability in DF emerging from a reconnection region in 3D simulations of sufficient domain size to track the evolution for several ion skin depths.

Recent Cluster multispacecraft  observations provide direct evidence for the presence of an interchange instability at DFs\citep{runov2012multipoint} at scales comparable with that observed in simulations~\citep{guzdar2010simple,lapenta2011self,vapirev2013formation}.

In the present paper, we consider the issue of the energetic consequences of the DF instability. We track the electromagnetic energy through the system and observe its balance, as described by the electromagnetic energy equation, and its flow described by the Poynting vector.

 The issue has been considered  from an observational angle in the recent work by \citet{hamrin2012role}. The energy exchange between plasma and fields is measured by computing directly $\bfJ \cdot \bfE$.  The measure is not easy, requiring the estimation of the current from the four Cluster spacecraft data \citep{hamrin2011energy}. Nevertheless, it has been done and the published results are used here to compare with our simulation data. 
 
Previous 2D studies have already investigated the energy exchanges. The energetics at the DF has been inferred from Cluster data by \citet{huang2012kinetic}, finding an energy transfer from the fields to the plasma in DF, in agreement with  2D simulation results \citep{sitnov2009dipolarization}. The 2D  simulations show a more intense electron contribution to the energy deposition near the x-point and a more intense ion deposition elsewhere~\citep{sitnov2011onset}. 

The extension to 3D can, and as shown below indeed does, explain several other features observed in data. 

First, the observations \citep{hamrin2012role} show that the energy exchange is exclusively from the fields to the plasma near the reconnection site, but as one considers Cluster crossings closer to the Earth the presence of energy exchanges in both directions is found. The condition where energy is going to the plasma from the fields is called a load, with circuit terminology. The opposite situation of energy being transferred by the particles to the field is called a generator. The presence of transfer of energy in both directions is shown also in MHD models~\citep{birn2005energy}, with loads concentrated in the mid-nigh region and generator regions located in the flanks where the cross-tail current is diverted
to the field-aligned currents of the substorm current wedge. The mechanism at play here is  different and due to the development of the DF instability.

Second, there is observational evidence for an important role of waves  generated at the DF \citep{marghitu2006experimental} where  generator regions  radiate electromagnetic energy. The suggestion is made that the energy goes into kinetic \alf waves.

The study for the  energy fluxes from Cluster observations shows that the dominant
component of the energy flux is ion enthalpy flux, with smaller contributions from the electron enthalpy and heat flux and the ion kinetic energy flux\citep{angelopoulos2002plasma,eastwood2013energy}. The Poynting flux is a minority contribution but it is not negligible, and in certain
parts of the ion diffusion region the Poynting flux in fact dominates\cite{eastwood2013energy}

The present description revisits these issues using 3D fully kinetic simulations, measuring directly the energy balance and reproducing many of the observational features outlined above. In particular, we observe that electromagnetic energy is converted to plasma energy at the reconnection site in a region very elongated along the separatrices. At the DF, we find the instability to convert about one order of magnitude more energy than at the reconnection site itself. But this energy is alternating between generator and load regions, as in the observations mentioned above. 

Significant regions of intense Poynting flux emerges from both regions of energy exchange. At the reconnection site, the energy flux is again of definite sign, but at the DF alternates in sign as the source causing it.

The remainder of the paper is organized as follows: Section 2 reports the details of the simulation approach followed and gives an overall view of he processes developing in the simulation. Section 3 considers the energy balance equation for the electromagnetic energy and investigates each term in turn. Section 4 provides a summary and an interpretation of the results of the energy analysis making a direct link with observations. The final overview is provided in Fig.~\ref{scenario} that summarizes pictorially the findings of the present investigation.

\section{Overview of the simulation}

We consider an initial Harris equilibrium:
\begin{equation}\label{initial}
\begin{array}{c}
\bfB =B_0 \tanh(y/\delta) \bfxhat + B_g \bfzhat\\
   \\
 p=p_b+p_0 {\rm sech}^2(y/\delta)
\end{array}
\end{equation}
with guide field: $B_g= B_0/10$ equal to one tenth of the maximum in plane field. The coordinates are chosen as: $x$ along the sheared component of the magnetic field (Earth-Sun direction in the Earth magnetosphere), $y$ in the direction of the gradients (north-south in the magnetosphere) and $z$ along the current and the guide field (dawn-dusk in the magnetotail).

The plasma of the Harris equilibrium is initially Maxwellian with a uniform drift that is prescribed by the force balance. A uniform background ($p_b$) is added in the form of a non-drifting Maxwellian at the same temperature of the main Harris plasma.

We consider the evolution from the same initial Harris current sheet of properties discussed in \citet{vapirev2013formation}: $v_{th,e}/c=.045$, $\delta/d_i=.5$ (where $d_i$ is defined based on $n_0$),  $T_i/T_e=5$, $m_i/m_e=256$, $p_b/p_0=.1$. The evolution is  then followed after adding a perturbation:
 \begin{equation}\label{perturb}
\delta A_z=A_{z0} \cos(2\pi x/L_\Delta)cos(\pi y/L_\Delta)e^{-(x^2+y^2)/\sigma^2},
 \end{equation}
of the vector potential, with $L_\Delta=10 \sigma$ and $\sigma=d_i/2$, where $d_i$ is the ion skin depth.

The study is conducted using a full kinetic treatment where both electrons and ions are treated as particles in the particle in cell approach. The code used is iPIC3D~\cite{iPIC3D} that is based on the implicit moment method~\cite{brackbill-forslund,lapenta2012particle}. This differs substantially from the standard explicit PIC method in that it removes the stability constraints of explicit methods and allows to focus the resolution only at the scales of interest, saving computational time. The simulation box has sizes $L_x/d_i=20$, $L_y/d_i=15$, $L_z/d_i=10$  discretized   in a grid $N_x=256$, $N_y=192$,$N_z=128$, with $N_p=5^3$ particles per cell using  a temporal resolution  of $\omega_{pi}\Delta t =0.125$. The grid is capable of resolving the background and (barely) the Harris electron skin depth $\Delta x/d_{eb}=0.4$, $\Delta x/d_{e0}=1.25$. These resolutions are lower than typical of explicit PIC, a feature allowed by the use of the implicit PIC method validated in many previous works~\cite{lapenta2012particle}. iPic3D uses a fairly standard normalization of all units, based on the speed of light, on the ion inertial length and ion plasma frequency. Below, results are shown in normalized units, but a few actual numbers are presented in the last discussion section.

Even though the implicit approach saves computational costs,  the simulation reported still requires massively parallel supercomputing, conducted using 1536 processors arranged in a 3D domain decomposition pattern of $16\times12\times8$ cores all of equal numbers of cells. Initially the particle are uniformly distributed with their weight chosen according to the local density. 

The typical evolution has been documented elsewhere~\cite{vapirev2013formation}. An extended x-line forms in consequence of the $z$-independent initial perturbation.  There, reconnection progresses in a 2D-like fashion where the topology retains the same configuration in each plane. This is direct consequence of the choice of perturbing the system in every plan $z$ in the same way. If a random perturbation is allowed the reconnection site acquires amuck more intricate topology~\cite{markidis-pop-2013}. 

The outflow from the reconnection region forms two fronts traveling along the plus and minus x direction. We refer to these front as downstream fronts (DF). The shorthand DF should not be confused with a dipolarization front. A dipolarization front proper develops only in the case of  realistic magnetotail configurations where the evolution of reconnection tends to restore a more dipolar configuration Earthward of the reconnection site. Here we start from an ideal Harris field, that is an appropriate approximation only sufficiently far from the Earth that the natural Earth dipole can be neglected. For this reason the DF should not confused with a dipolarization front.

A previous study of the evolution of the DF~\cite{vapirev2013formation} in 3D demonstrates the onset of a secondary instability at the front that leads to its rippling and formation of fingers of plasma that interact and merge.

\begin{figure}
\includegraphics[width=\columnwidth]{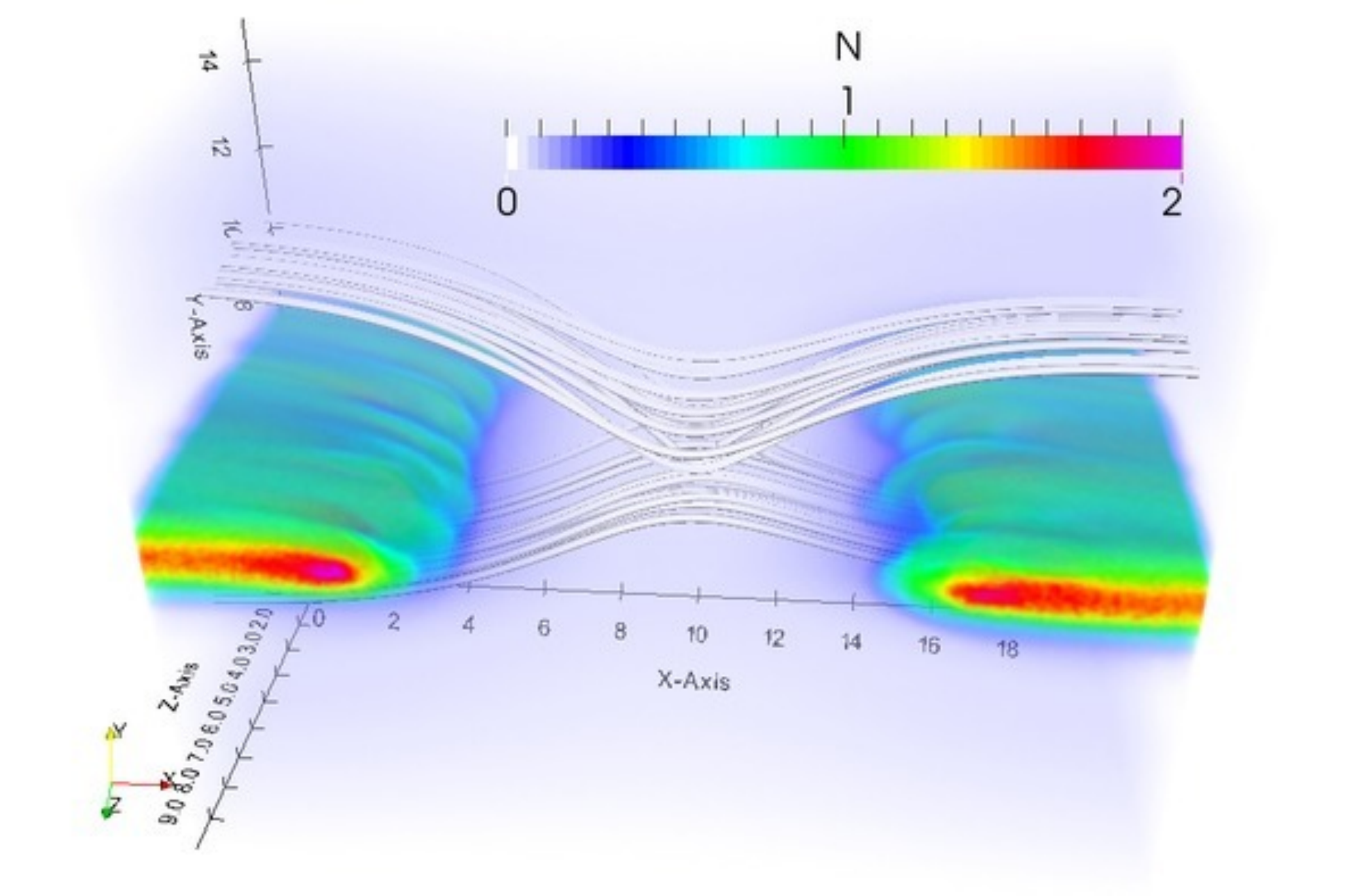}
\caption{Volume rendering (fog-like) of the ion density, $n/n_0$. To guide the eye selected portions of  field lines are reported, emerging from a sphere of radius $R/d_i=2$ at the center of the box. In this and all plots below, the time corresponds to cycle 13000, or time $\omega_{ci}t=15.763$ in run catalogued tred54 in the MMSIDS University of Colorado server.}
\label{rho}
\end{figure} 

Figure \ref{rho} shows the two DF traveling leftward and rightward and forming ripples that subsequently continue to grow and intensify.   The process is reminiscent of a type of Raleigh-Taylor  instability. However, the true nature of this instability in kinetic theory must take into consideration the presence of  density gradients leading to kinetic drift instabilities in the lower hybrid range~\citep{divin2013dipolarization}. 

 \begin{figure}
\includegraphics[width=\columnwidth]{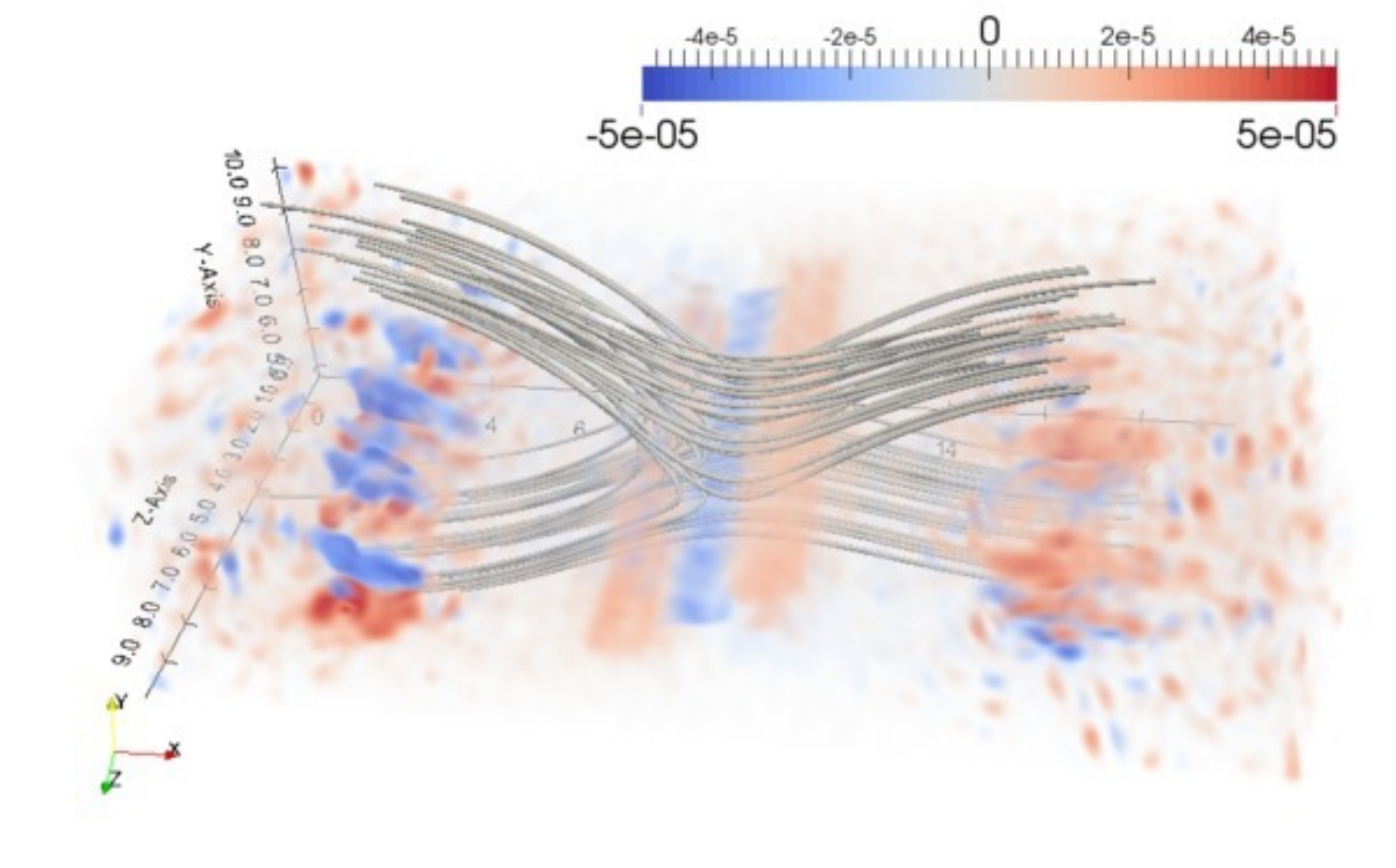}
\caption{Volume rendering of the normalized parallel electric field, $eE_{||}/cm_i\omega_{pi}$. To guide the eye, selected portions of field lines emerging from a sphere of radius $R/d_i=2$ are reported at the center of the box.}
\label{epar}
\end{figure} 
The electric field is affected by the development of the secondary DF instability. The parallel electric field (see Fig.~\ref{epar}) presents two types of regions: one in the proximity of the x-line and one at the two DFs. Here we define parallel, as customary, with respect to the local magnetic field direction.

The former, like the topology, retains the $z$-independence of the initial state. Close to the central x-line, the magnetic field is primarily directed along $z$ because the in $(x,y)$-plane field vanishes at the x-line. This is the reconnection electric field, electromagnetic in nature and negative as required to have the correct sign for $\partial A_z/\partial t$ that gives reconnection. 

Further away from the center the parallel electric field shows how even the modest guide field used of $B_g=0.1B_0$ breaks the symmetry~\citep{goldman2011jet}. Two separatrices (upper-left and lower-right) tend to develop more electron holes~\citep{mmspop,lapenta2011bipolar}. The other two present a more intense current. The electron jet from the reconnection site is deflected and tilted towards these latter two set of separatrices~\citep{goldman2011jet,le2013regimes} and the parallel electric field reverses sign, being negative in the center and positive at the ends of the jet~\citep{zenitani2011new} (note the opposite sign due to the different choice of axis when compared with \citet{zenitani2011new}).  These  process  develop just like in 2D studies and have been studied extensively in the past~\cite{kleva,ricciguide,mmspop,lapenta2011bipolar} and have been reviewed recently by Shay and Drake in Ref.~\cite{birn-priest}. 

More innovative is the other type of region, that at the two DFs. There the $z$-independence is broken, alternating positive and negative. This electric field is caused by the DF instability. The association of interchange-like or drift-like instabilities with a parallel electric field is a confirmation of the density gradient-driven nature of this instability.

 \begin{figure}
\includegraphics[width=\columnwidth]{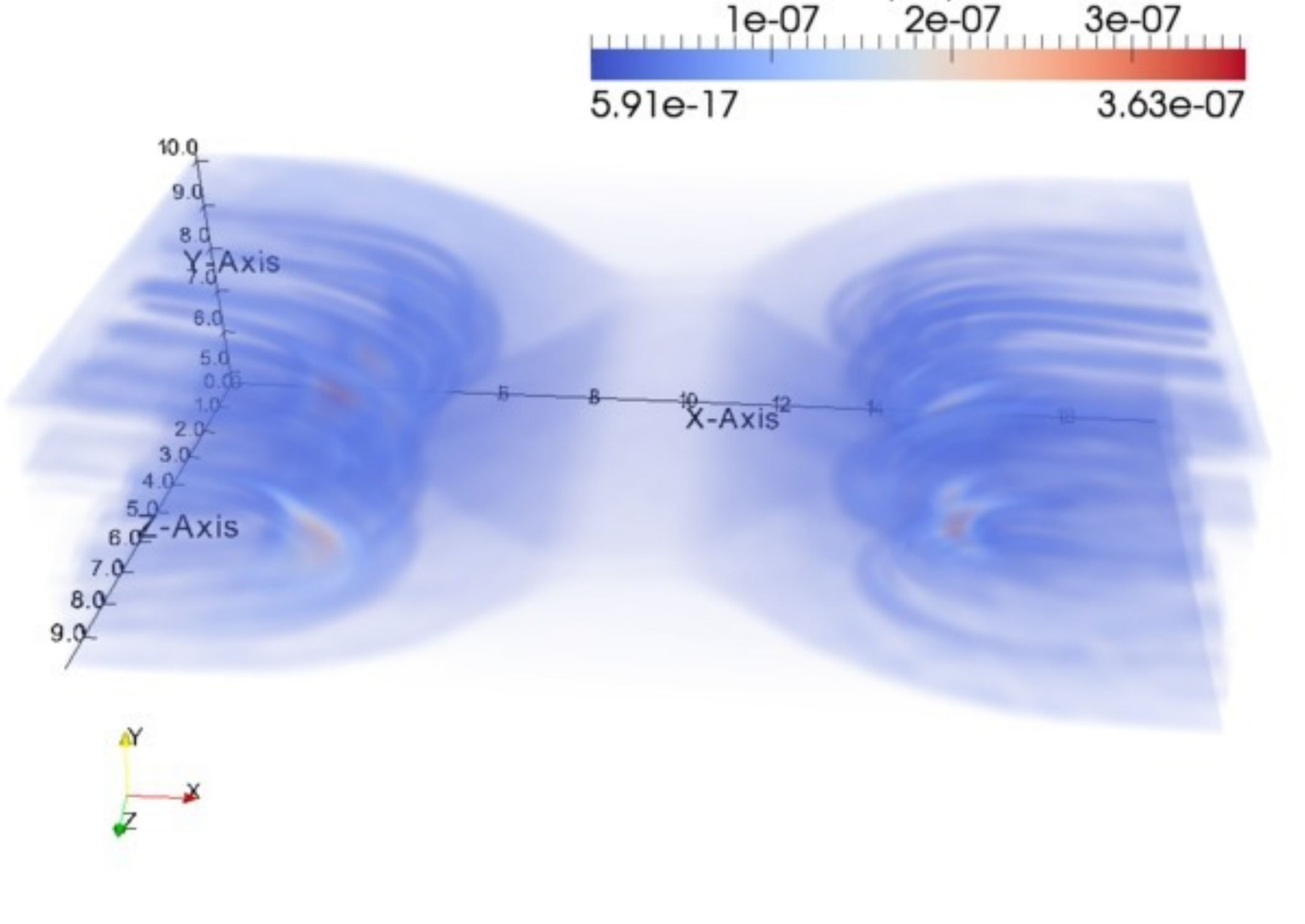}
\caption{Volume rendering of the magnitude of the magnitude of the perpendicular component of the electric field, $e\bfE_{\perp}/cm_i\omega_{pi}$. }
\label{eperp}
\end{figure}

The perpendicular electric field also presents two types of regions (see Fig.~\ref{eperp}).  
The first extends along the four separatrix surfaces and is caused by the Hall electric field, primarily electrostatic in nature and also fundamentally linked with kinetic reconnection, just like in 2D. The other structure emerges from the DF, nested inside the  perpendicular field at the separatrices. Both structures relative to the perpendicular field extend far from their origin. 

The field at the separatrices here extends to the boundary of the simulation. Previous 2D studies have uncovered that this field is associated with a Poynting flux that propagates at superalfv\'enic speeds~\cite{shay11,lapenta2013propagation}. Similarly the DF instability is also the source of a perpendicular field also extending far from the DF itself. The domain size can only probe the extension of the field to relatively small distances, but the analogy with the 2D results invite future studies to investigate the possibility that this field might expand for several Earth radii towards the Earth.

\section{Electromagnetic Energy Budget}

The focus of the present work is on the energetic consequences of the DF instability. In particular, on the electromagnetic part of the energy balance. We measure the electromagnetic energy  processed by the instability and show that it is largely dominant over any other energetic process in the system, exceeding by an order of magnitude event the energy released by the reconnection process itself. Reconnection is the primary cause but it is just a spark that ignites the real fire represented by the instability in the DF. We measure in the DF an energy exchange one order of magnitude larger than the energy released in reconnection. Furthermore, while reconnection is  a process primarily converting magnetic energy to kinetic energy, where the flow of energy is almost uniquely positive, from the field to the plasma, the DF instability presents regions of both signs with energy flowing both from the plasma to the fields as well as the viceversa.

The balance of  electromagnetic energy in any system is given by
\begin{equation}
\frac{1}{2}\frac{\partial }{\partial t}\left( \epsilon_0 E^2 +\frac{1}{\mu_0}B^2 \right) = -\bfE \cdot \bfJ -  \nabla \cdot \bfS
\label{energy-balance}
\end{equation}
where $\bfS=\bfE \times \bfB/\mu_0$ is the Poynting vector.  Equation (\ref{energy-balance}) is valid in vacuum (where the current is zero of course) as well as in any medium, plasma included, and it is based just on the properties of the Maxwell equation.  The change in local electromagnetic energy is determined by two factors. First, the exchange of energy with the plasma ($\bfE \cdot \bfJ$): where a positive value means that energy is going from the field to the plasma (the minus sign in eq. (\ref{energy-balance}) correspondingly subtracts energy from the fields). Second, the divergence of the Poynting flux provides a radiative process whereby electromagnetic energy can be propagated via waves. In vacuum these can only be light waves propagating at the speed of light, but in a medium, any other wave or process leading to a divergence of the Poynting flux is also a viable mechanism to transfer electromagnetic energy.
In the case of reconnection, recently the role of kinetic \alf waves (KAW) as energy carrier via the Poynting vector term has attracted special attention \cite{shay11, lapenta2013propagation}. 

Each term of the equation has been analysed for the simulation described above. Let us analyze them in turn, starting from the energy exchange between plasma and field, followed by the Poynting term and finally the change in local energy content.

\subsection{Energy exchange between plasma species and fields}
To appreciate the energy exchange between plasma and the electric field, two types of visualizations need to be considered together.

\begin{figure}
\includegraphics[width=\columnwidth]{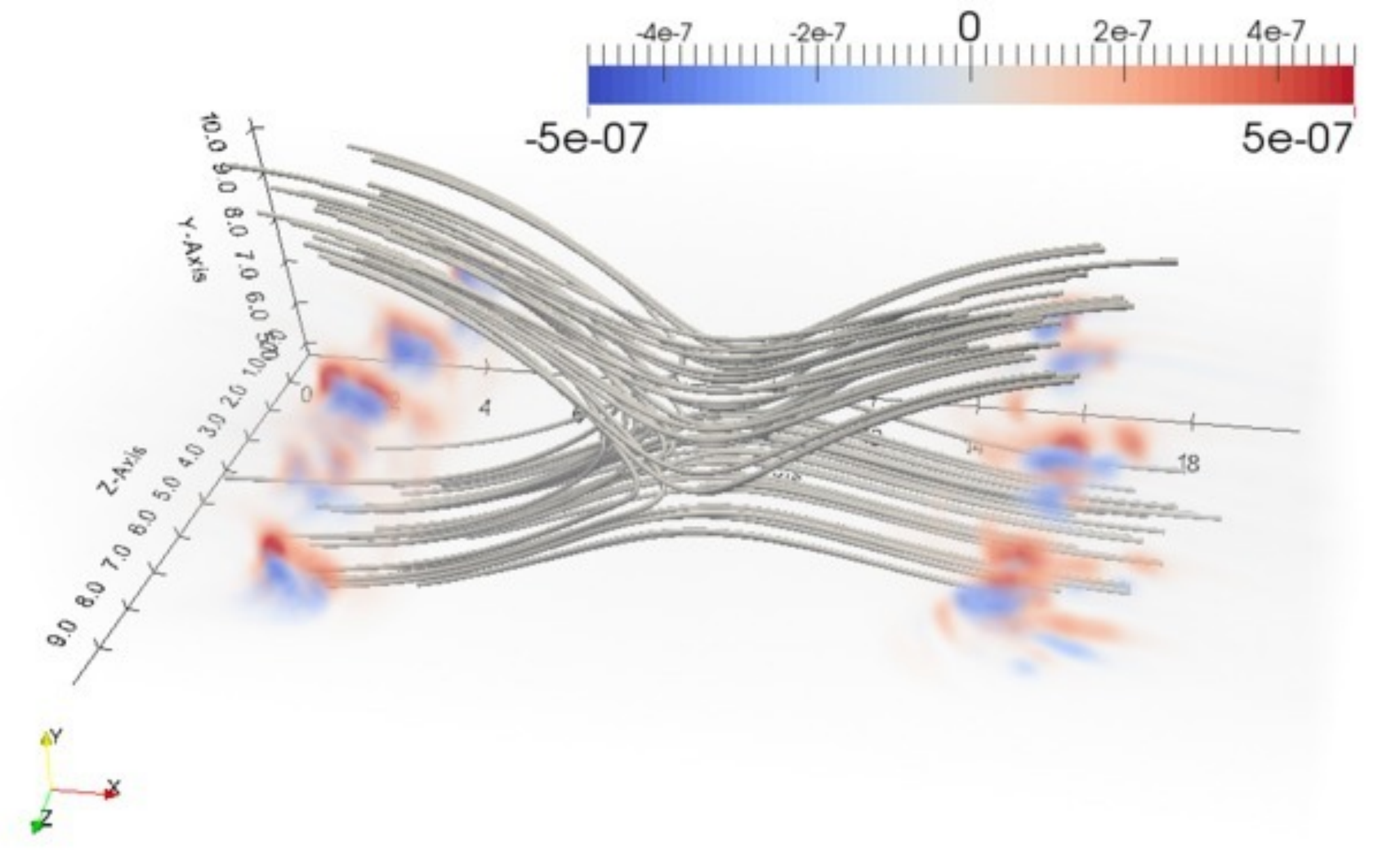}
\caption{Volume rendering of the normalized energy exchange term, $\bfJ_i \cdot \bfE \;d_i/n_0c^2m_i$, for ions. Values close to zero in light gray are made transparent by properly choosing the transfer function. To guide the eye, selected portions of field lines emerging from a sphere of radius $R/d_i=2$ are reported at the center of the box.}
\label{jidotE}
\end{figure} 

\begin{figure}
\includegraphics[width=\columnwidth]{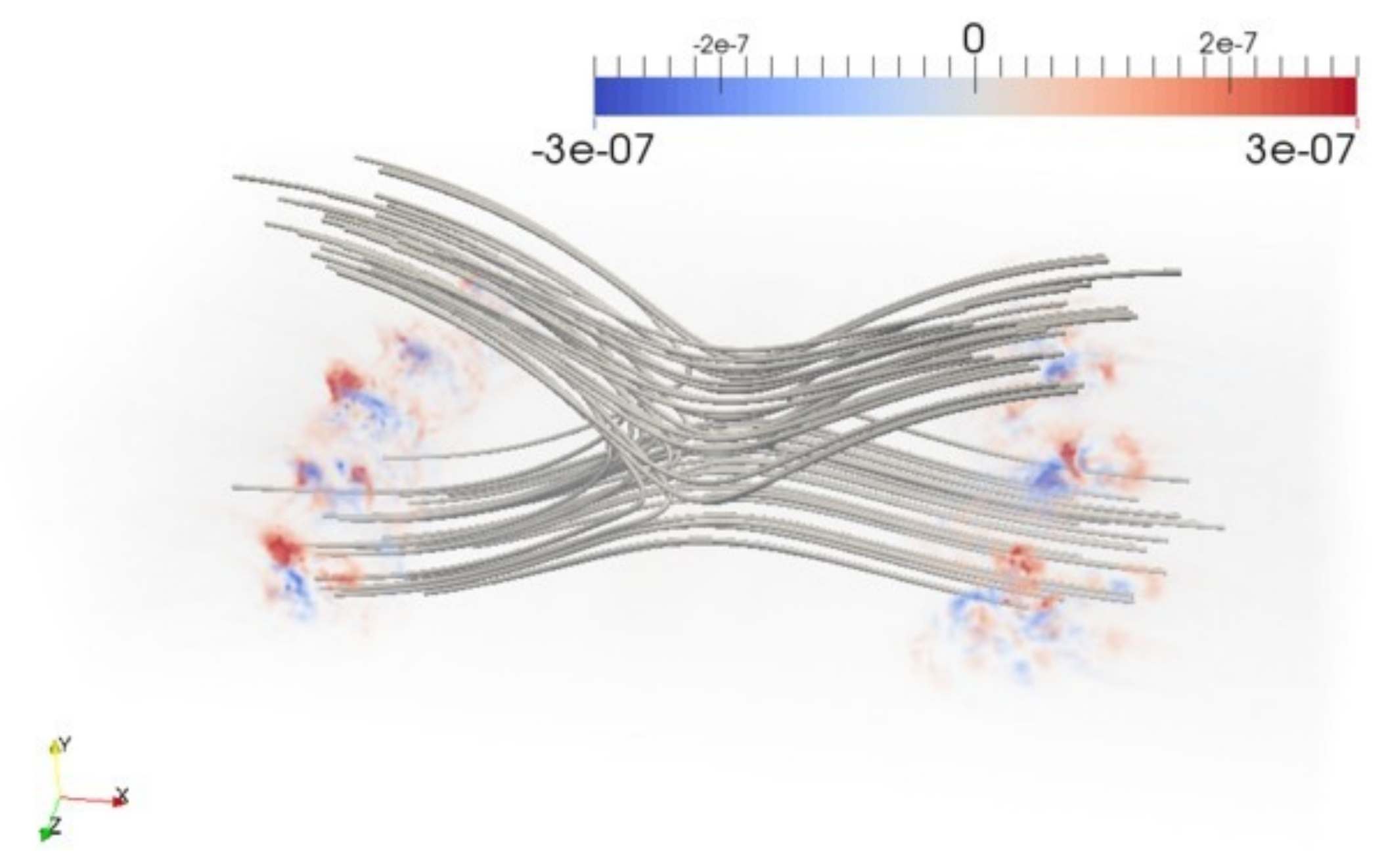}
\caption{Volume rendering of $\bfJ_e \cdot  \bfE\;d_i/n_0c^2m_i$ for electrons. Values close to zero in light gray are made transparent by properly choosing the transfer function. To guide the eye, selected portions of field lines emerging from a sphere of radius $R/d_i=2$ are reported at the center of the box.}
\label{jedotE}
\end{figure}

First, the local value of $\bfJ_s \cdot  \bfE$, for each species $s$, is visualized with volume (fog-like) rendering in 3D. Figure \ref{jidotE} and Fig.~\ref{jedotE} report the value, respectively, for  ions and electrons. Clearly the energy exchange is localized in the DFs and it has an oscillating nature along $z$ with positive and negative regions alternating. A positive $\bfJ_s \cdot  \bfE$ corresponds to a situation where the electric field transfers its energy to the particles (in electric circuit language this corresponds to a load). Conversely a negative value is present in regions where the energy of plasma is transferred to the field (in circuits this corresponds to a generator). In the DF, generator and loads are alternating along $z$ following the rippling of the DF caused by the instability. 

Barely visible on paper is a small exchange also at the separatrices and near the x-line.  To discern these other components, we consider a second type of visualization approach. The value of $\bfJ_s \cdot  \bfE$ is observed to oscillate along $z$, suggesting to consider the average along $z$. We define then for a generic quantity $\Psi$, the z-average as:
 \begin{equation}
<\Psi>=\frac{1}{L_z}\int_0^{L_z}\Psi dz.
\end{equation}

\begin{figure}
\includegraphics[width=\columnwidth]{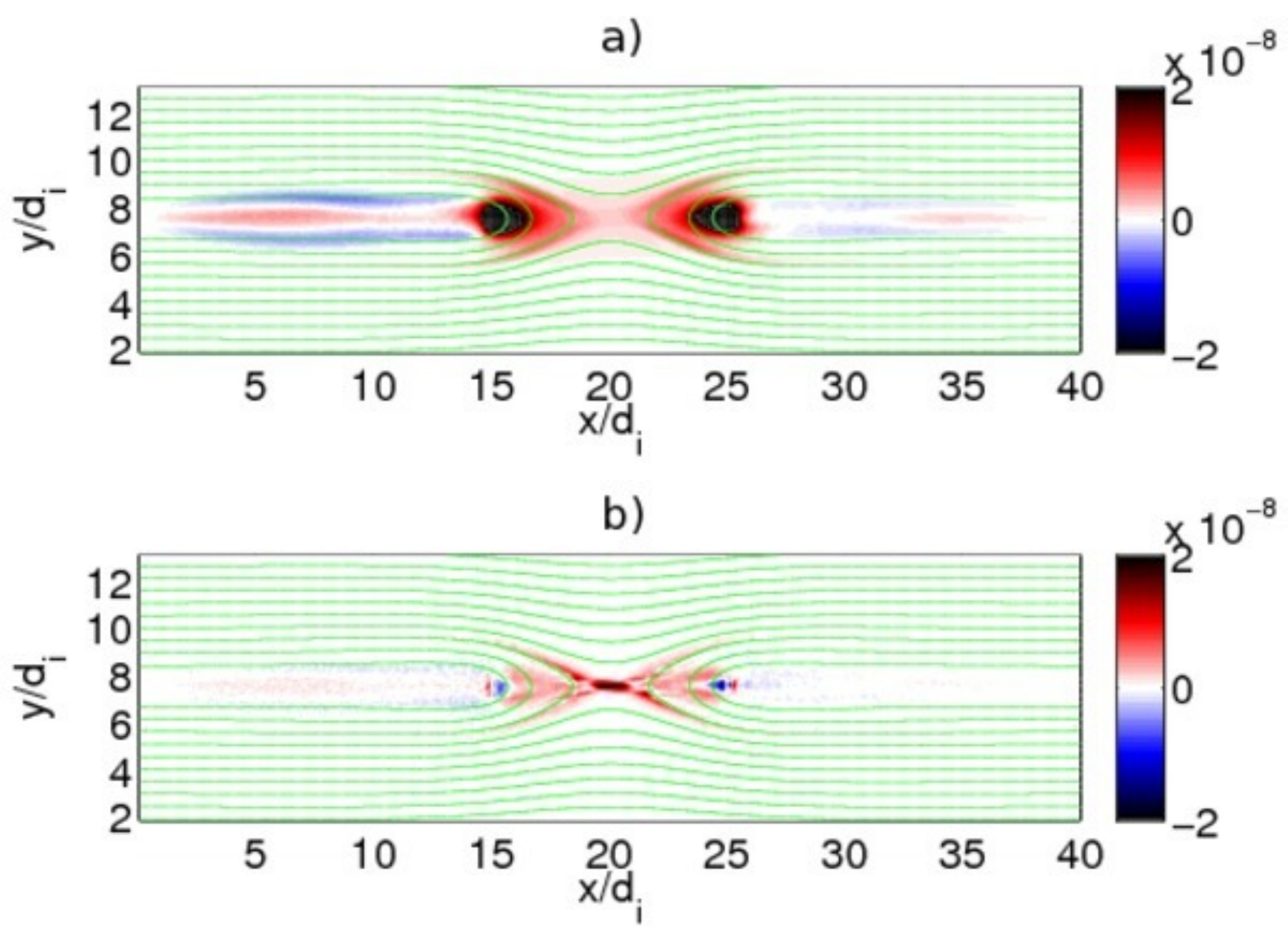}
\centering
\caption{Average along $z$ of the normalized energy exchange term in the energy balance equation: a) $\langle\bfJ_i \cdot \bfE\rangle_z\;d_i/n_0c^2m_i$ (top); b) $\langle\bfJ_e \cdot \bfE\rangle_z\;d_i/n_0c^2m_i $ (bottom). }
\label{avgJdotE}
\end{figure}

Figure \ref{avgJdotE} shows $\langle \bfJ_s \cdot  \bfE\rangle_z$ for electrons and ions, where the symbol $\langle\;\rangle_z$ is used to indicate the average along $z$. 

Two evidences jump at the inspection. 

First, the color scale. The volume rendering of the full $\bfJ_s \cdot  \bfE$ is more than one order of magnitude larger than the z-averages at $5 \cdot 10^{-7}$ versus $ 2\cdot 10^{-8}$. The energy exchange in the DF takes place nearly in as many loads as generator regions, nearly averaging out to null. For the electrons the average in the DF is very small, much smaller than the energy exchange happening along the separatrices and near the x-line. For the ions there is a more sizable residual energy gain. 

Average is a dangerous concept as an hungry person would readily testify watching someone eating two servings.  The fact that on average as many particles gain energy as others loose energy does not mean that no energy is being exchanged. But it does mean that on average the overall plasma gains as much at it looses. 

Second, at the x-line and at the separatrices the electrons acquire most of their energy gain, with a positive value of $\bfJ_e \cdot  \bfE$. The ions instead are not visibly affected there, instead gaining most of their average energy in the DF. 

The behavior of the $z$ averages is consistent with  previous findings  \citep{weigang-2008-evolutions,sitnov2009dipolarization}. The reconnection electric field $E_z$ is observed to move its peak during the simulation. Initially it is centered at the x-line, but as the DF form, the peak splits and moves outward  remaining focused at the DF \citep{weigang-2008-evolutions}. There the energy exchange is maximum for the ions \citep{sitnov2009dipolarization}. For the electrons instead, the main energy exchange remains more focused on the central electron diffusion region and along the separatrices. 

The energy deposition to the particle is represented by two terms: particle heating and flow. The former results in actual temperature increase while the latter produces mean flows without increasing the local temperature. Future work will investigate the energy balance equation for the plasma. The focus here is only on the electromagnetic energy balance.

\subsection{Poynting vector}
The next term in the electromagnetic energy balance equation is that expressing the flow of electromagnetic energy expressed by the Poynting flux. The Poynting vector itself represents the momentum associated with electromagnetic fields, and its divergence expresses the changes in the contact of energy due to the transmission of waves. Obviously these can be electromagnetic waves, that in a medium need not travel at the speed $c$ in vacuum. These can in fact be standing waves as well.  

\begin{figure}
\includegraphics[width=\columnwidth]{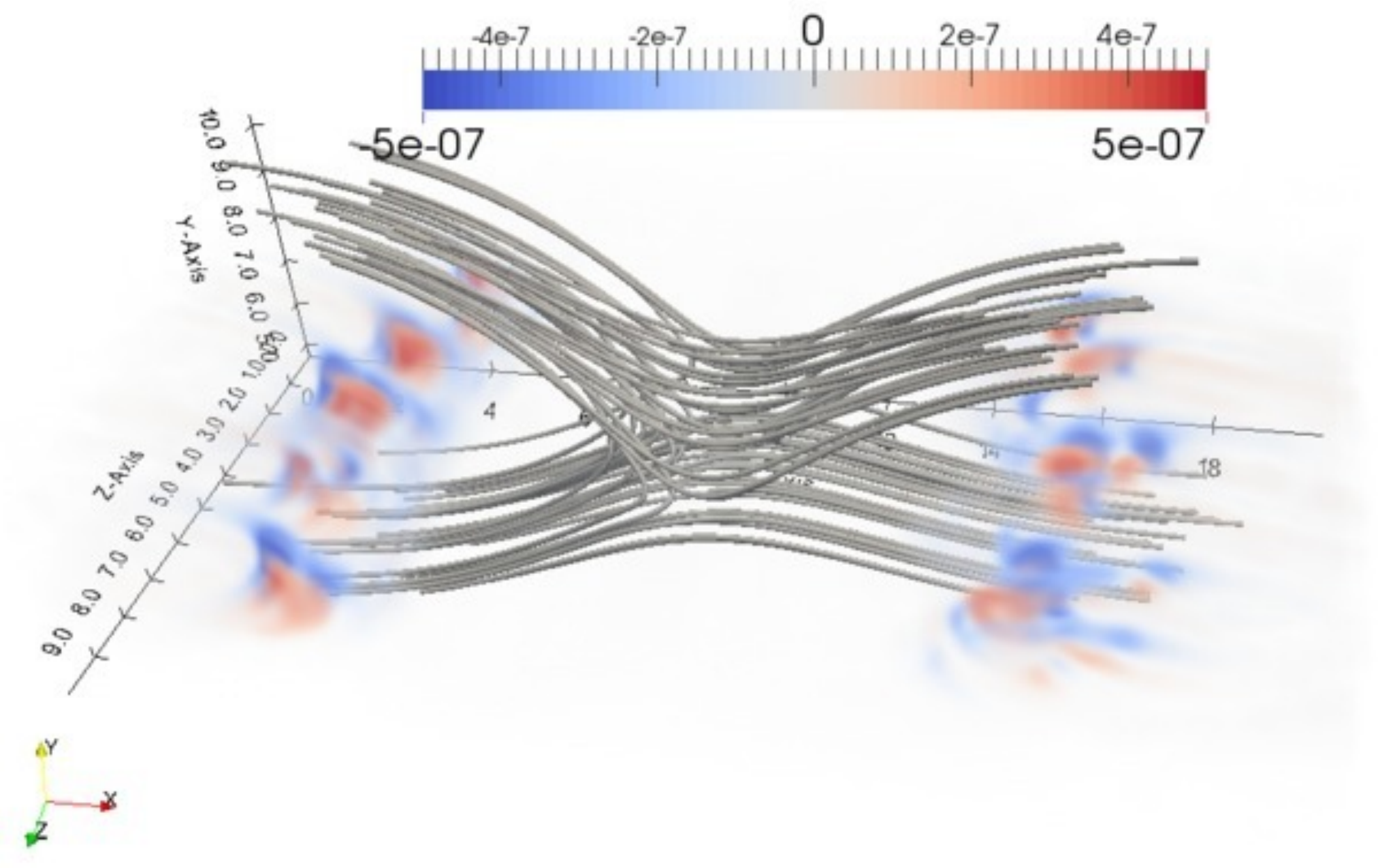} 
\centering
\caption{Volume rendering of the normalized divergence of the Poynting flux, $e^2\nabla \cdot \bfS /m_i^2\omega_{pi}^3$.  Values close to zero in light gray are made transparent by properly choosing the transfer function. To guide the eye, selected portions of field lines emerging from a sphere of radius $R/d_i=2$ are reported at the center of the box.}
\label{divS}
\end{figure}

Figure \ref{divS} shows the divergence of the Poynting flux observed in the run at the same time considered above. Again the energy flux term is by far the strongest in the DF and again it is oscillating between regions of net gain and net loss. 

\begin{figure}
\includegraphics[width=\columnwidth]{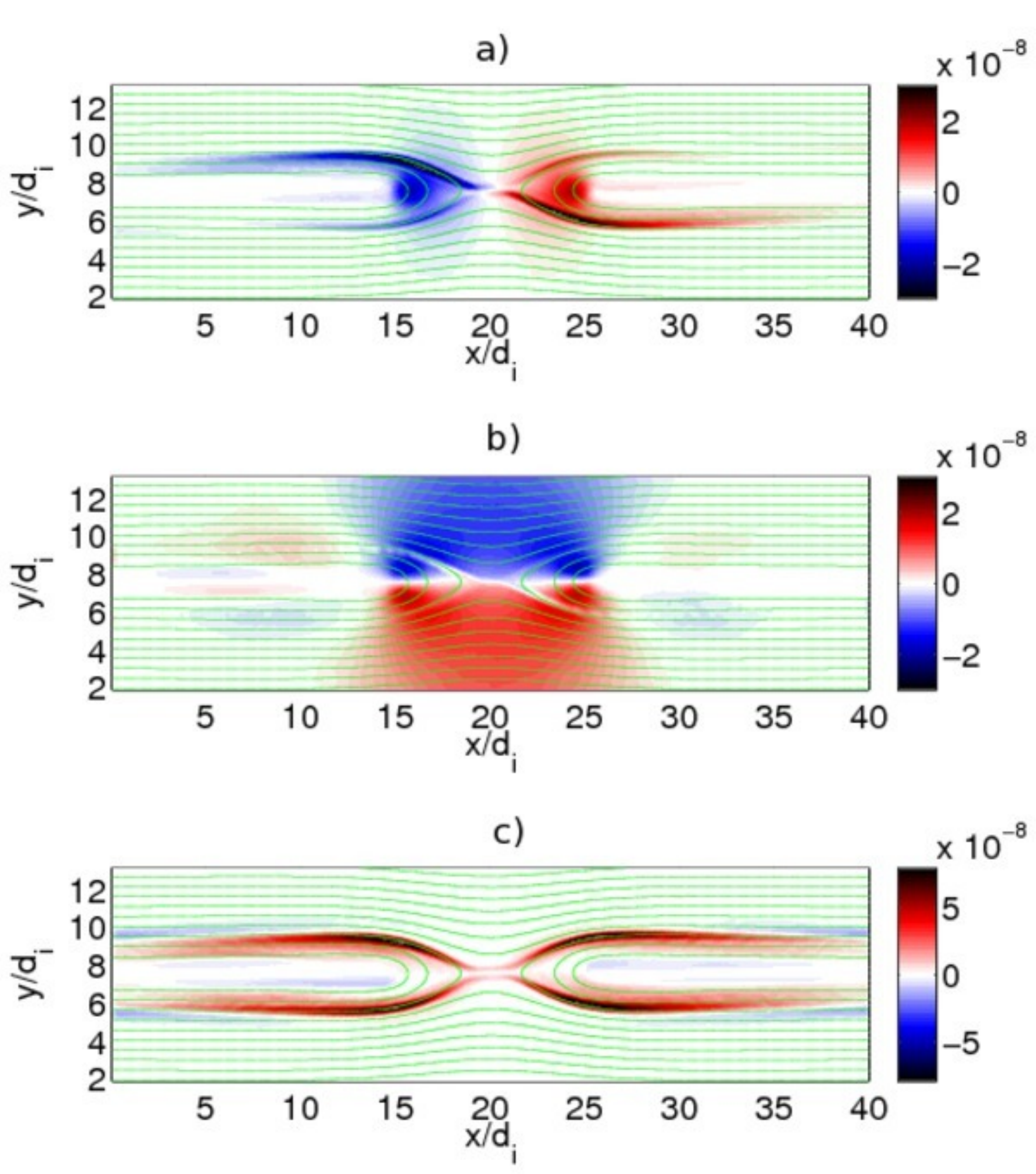} 
\centering
\caption{Normalized Poynting flux. On the left,  contribution form average fields is given: $S_{\rm mean}=e^2\langle\bfE\rangle_z\times\langle\bfB\rangle_z/\mu_0m_i^2\omega_{pi}^2$. From top to bottom: a) $x$-component, b) $y$-component, c)   $z$-component.}
\label{AVGS}
\end{figure}

Recently, the presence of a strong energy flux traveling along the separatrices has been shown to travel at superAlfv\'enic speeds \cite{shay11, lapenta2013propagation} in 2D simulations. Figure \ref{AVGS} shows the Poynting flux due to the average fields $\langle\bfE\rangle_z\times\langle\bfB\rangle_z$. The average fields reproduce a contribution similar to the 2D results and localized at the separatrices.

\begin{figure}
\includegraphics[width=\columnwidth]{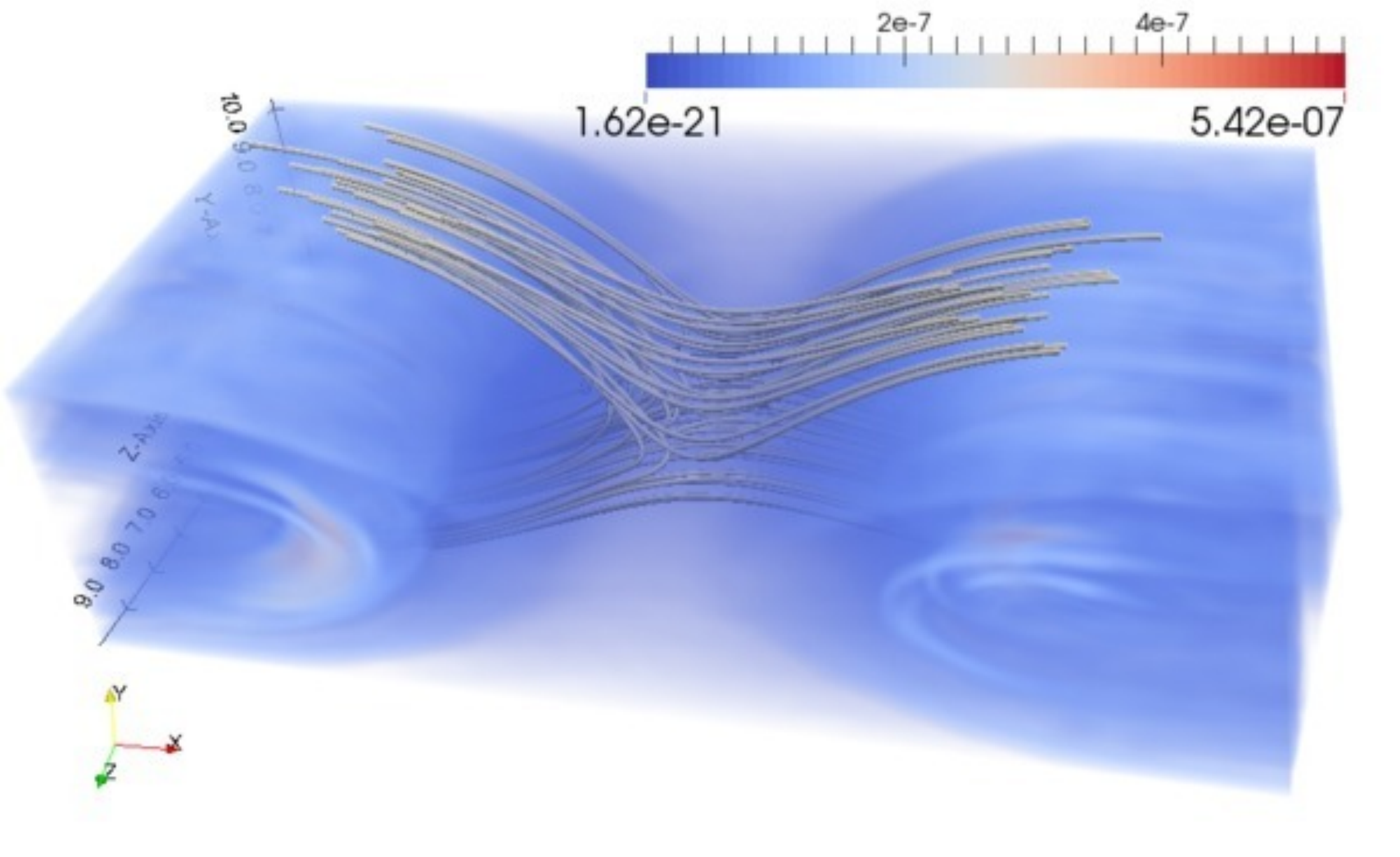} 
\centering
\caption{Volume rendering of the magnitude of the normalized Poynting flux. To guide the eye, selected portions of field lines emerging from a sphere of radius $R/d_i=2$ are reported at the center of the box.}
\label{poynting}
\end{figure}

However, a 3D volume rendering (see Fig.~\ref{poynting}) of the full Poynting flux $\bfS$ shows again a great contribution coming from the DF. In 3D then, two  flows of electromagnetic energy are generated. One progresses along the separatrices and is caused by the Hall physics generated by the process of reconnection, the other springs out of the energy exchange between particles and field at the DF. Locally this second contribution is stronger, but as it propagated out it becomes comparable to that at the separatrices. 

Another great difference is present between the two Poynting flux contributions. That on the separatrices is independent of $z$ and present a 2D-like configuration. The other originating from the DF is alternating in sign just like its source and when the averages are made along $z$ its contribution tends to average out and disappears from the average means. 

The fluctuating fields instead produce a Poynting flux propagating from the DF, closer to the neutral plane (the plane where the initial in plane magnetic field reverses sign, $y=L_y/2$). 

The large energy exchanges developing at the DF are also a source of intense electromagnetic energy fluxes expressed by the Poynting flux.

\subsection{Change in energy density}

\begin{figure}
\includegraphics[width=\columnwidth]{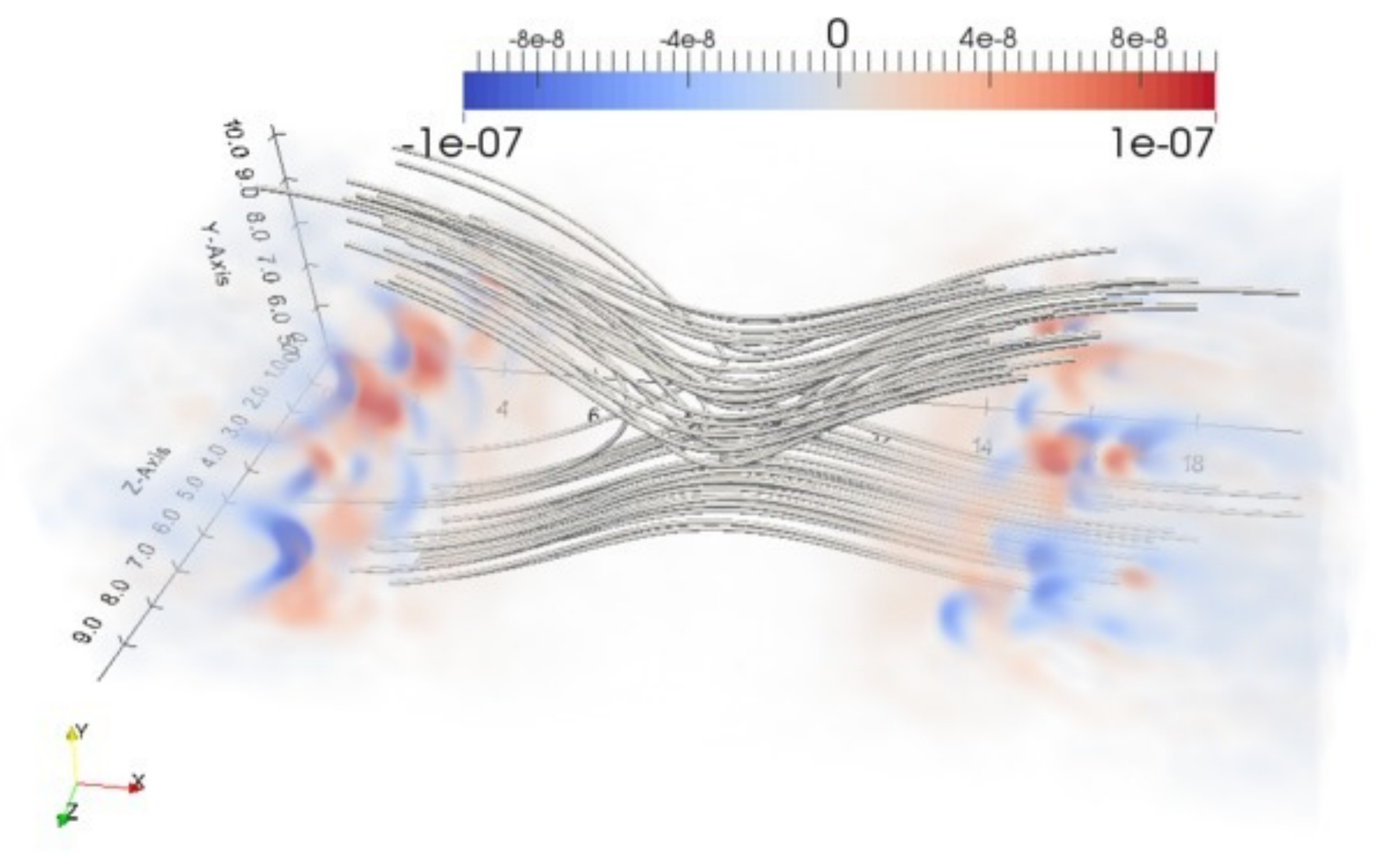}
\centering
\caption{ Volume rendering of $\omega_{pi}^{-1}\partial W_B/\partial t$, where $W_B=e^2B^2/2\mu_0m_i^2\omega_{pi}^2$. To guide the eye, selected portions of field lines emerging from a sphere of radius $R/d_i=2$ are reported at the center of the box.}
\label{db2dt}
\end{figure}

The last term of the balance is the local content of the electromagnetic energy density. The vast majority of the energy is held by the magnetic field, with the electric field energy budget being negligible by comparison. This is of course a consequence of the fact that the changes are here happening at a speed vastly inferior to the speed of light. So the only change to be concerned about is the change in magnetic energy content, shown in Fig. \ref{db2dt}. This term also is focused on the DF and there showing both positive and negative net changes. 

Unlike the other two terms that were more localized at the edge of the DF towards the central x-line, the change in magnetic energy density is more spread in the full width of the DF. 

\section{Energetics: An emerging scenario}
\begin{figure}
\includegraphics[width=\columnwidth]{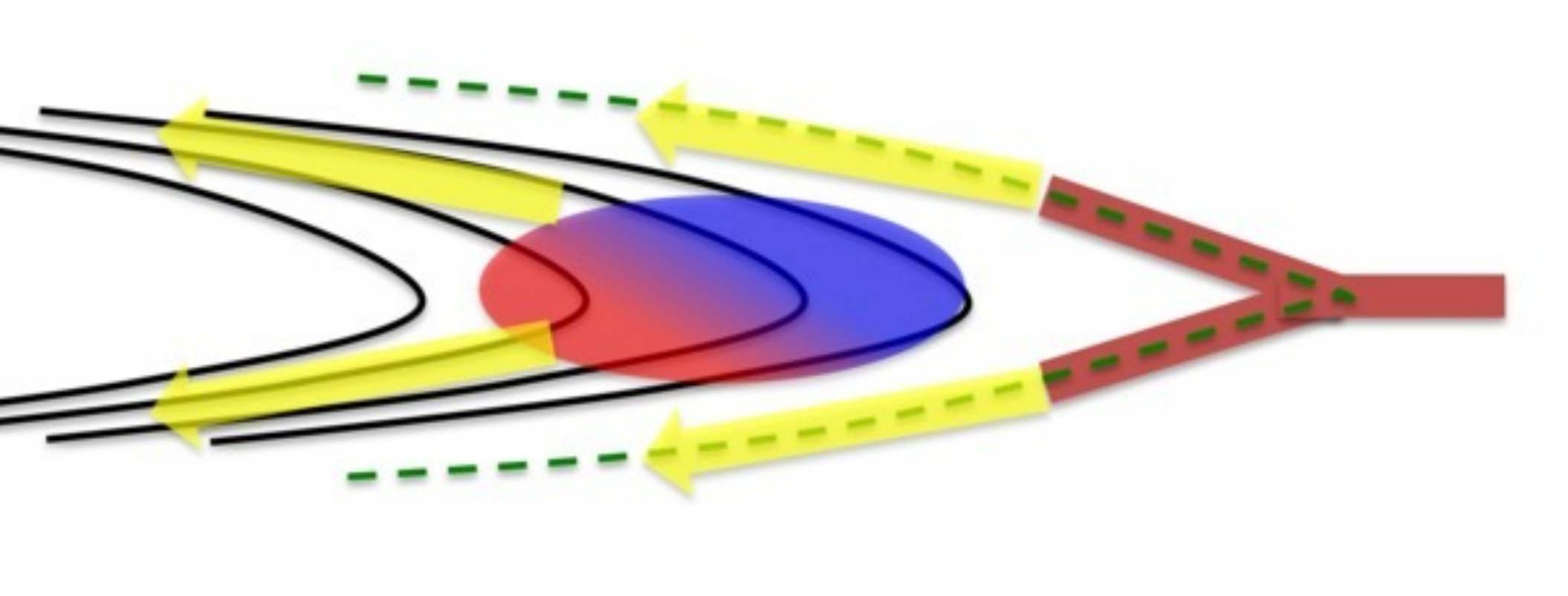}
\caption{Summary  cartoon of the electromagnetic energy balance. The reconnected field lines are shown by black solid lines, the separatices as dashed green line. The energy exchange $\bfJ \cdot \bfE$ is focused on the vicinity of the x-line and stretching along the separatrices and in the downstream front (DF) at the nose of the reconnected lines. In the x-line vicinity the energy exchange is positive (marked in red) going from the field to the plasma. At the DF, it is alternating in sign but more of an order of magnitude larger. From the two energy exchange regions a Poynting flux emerges (marked in yellow) traveling downstream away from the reconnection region. Another similar structure develops on the right side of the x-line but it is not shown here. }
\label{scenario}
\end{figure}

The results obtained above can be summarized in the cartoon scenario shown in Fig.~\ref{scenario}. The exchange of energy between plasma and fields is  concentrated at the x-line in a region stretching along the separatrices. There the energy is being transferred from the fields to the particles. A second and more than an order of magnitude more intense energy exchange happens at the DF and is mediated by the instability developing at the DF. 

This second energy exchange region is absent in 2D simulations and even in 3D it disappears when an average along $z$ is made. In this region the energy exchange oscillates in sign with load regions where  electromagnetic energy is deposited to the particles balanced by generator regions where particle energy is transferred to the fields. The mechanism for this energy transfer involves both the parallel and perpendicular electric fields and is linked to the density-gradient driven instability at the front. 

From both energy exchange regions Poynting fluxes emerge traveling downstream away from the region of reconnection. On the separatrices, the Poynting flux is generated by the Hall fields and has been shown to travel at superalfv\'enic speeds~\cite{shay11, lapenta2013propagation}. The Poynting flux generated at the DF, instead, is due to the DF instability. The two fluxes are very different in one aspect: the separatrix flux is of definite sign and independent of $z$, the flux emerging from the DF instability is alternating in sign and dependent on $z$ in the same way as the source generating it (the DF instability). The Poynting flux observed here can only travel the relatively short distances allowed by the computational box used. Larger boxes will be needed to determine the extent of the range of the Poynting flux generated.

The energy associated with this flux is very substantial. Recently, \citet{shay11} computed the entity of the separatrix Poynting flux to be significant and in fact strong enough to impact auroral physics. The second Poynting flux emerging from the DF and cued by the DF instability is of comparable size.  

To make a more direct comparison with observations, it is useful to put the results of the simulations above, so far reported in dimensionless ratios, in physical SI units. To fix the ideas we consider a typical reference situation where the  plasma density  is $n_0=0.1\unit{cm}^{-3}$ and $B_0=20\unit{nT}$. At the chosen thermal speed used in the simulations, physical electrons would have a speed of $v_{th,e}=0.045$ when the ion temperature is $T_i=5\unit{keV}$ and the electrons $T_e=1\unit{keV}$. Since a mass ratio of 256 is used, this means we are using unphysically light ions. Using these numbers, the physical value corresponding to a code value of $(\bfJ_s \cdot \bfE)_{\rm code}=10^{-8}$ is  $\bfJ_s \cdot \bfE=1.86 \unit{pW/m^3}$.  The numbers observed in the simulations then range in the order $\unit{pW/m^3}$ in the x-line  and  ten times more at the peak in the DF. These are numbers consistent with those reported in the observational evidence~\citep{marghitu2006experimental} and in MHD-based investigations of the global magnetosphereic energy cycle~\citep{birn2005energy}.

For those same parameters, a Poynting flux 
of $\bfS_{\rm code}=10^{-7}$ in the simulations corresponds to $\bfS=8 \cdot 10^{-4}\unit{W/m^2}$, a value consistent with the recent analysis by \citet{shay11, lapenta2013propagation} and the observational evidence reported in \citet{eastwood2013energy}.

Besides the order of magnitude of the energy exchange and of the Poynting flux, the results from the simulations reported above present other important similarities with observed data. The most comprehensive summary of recent observations is reported in \citet{hamrin2012role}. The analysis of energy transformations in the tail has focused on measuring the Poynting flux, particle flux and the work between fields and particles ($\bfE \cdot \bfJ$). 

The result is that both generator  and load regions are indeed observed. Generator regions are less common and concentrated closer to Earth, significantly Earthward of   the near Earth neutral line (NENL) where reconnection develops. Load regions are concentrated nearer the NENL~\citep{hamrin2011energy,hamrin2012role}. These observational findings are clearly in agreement with the results above that also show an univocally load region near the x-line and an alternating generator and load zone in the DF in conjunction with the instability.

A number of other important properties have been observed, suggestive of further agreement with the simulation results above. First,  generator regions  tend to be off the center of the current and more towards the edge of the sheet bordering the lobe \citep{marghitu2006experimental, hamrin2009occurrence}.  Second, the $GSM-y$ (dawn-dusk) component (corresponding to the simulation $z$ coordinate)  is generally dominant in load regions, but generator regions appear more complex and related to wavy structures where not only GSM-$y$ dominates but other components become important as well\citep{hamrin2006observations},  an indication of a higher level of fluctuations in correspondence with generator regions \citep{hamrin2009occurrence}. This is obviously a possible support to the finding above that generator regions alternate with load regions in the DF as a consequence of the instability developing there. 

The presence of the Poynting flux is also supported by observations. \citet{angelopoulos2002plasma} note that observations demonstrate that in bursty bulk flows (BBF) energy is primarily carried by particle energy fluxes but the Poynting flux is a significant minority contribution. A more recent analysis of Cluster data supports this conclusion \citep{eastwood2013energy}. The energy is found to be dissipated in large part before reaching the ionosphere of the Earth suggesting that it might be dissipated into \alf or kinetic \alf waves at the dipolarization front \citep{marghitu2006experimental} 

Based on the observational conclusions summarised above, the possible  scenario illustrated in Fig.~\ref{scenario} emerges in agreement with the results reported here.

\section*{Acknowledgments}
The present work is supported by the NASA MMS Grant
NNX08AO84G. Additional support for the KULeuven
team is provided by the European Commission Seventh Framework Programme (FP7/2007-2013)
under the grant agreement no. 263340 (SWIFF project,
www.swiff.eu), by the Onderzoekfonds KU Leuven (Research Fund KU Leuven) and 
by the Interuniversity Attraction Poles Programme of the Belgian Science Policy Office (IAP P7/08 CHARM).

The simulations were conducted on the Pleiades supercomputer of the NASA Advanced Supercomputing Division (NAS), on the Discover supercomputer of the NASA Center for Climate Simulation (NCCS) and on the TGCC Curie supercomputer thanks to the PRACE research infrastructure grant SWEET.

\end{document}